\pdfoutput=1

\documentclass[11pt]{article}
\usepackage[final]{acl}

\usepackage{times}
\usepackage{latexsym}

\usepackage[T1]{fontenc}

\usepackage[utf8]{inputenc}

\usepackage{microtype}

\usepackage{inconsolata}

\usepackage{graphicx}
\usepackage{xcolor}
\usepackage[skip=2pt]{caption}
\usepackage[hang,flushmargin]{footmisc}
\usepackage{amssymb}
\usepackage{booktabs}
\usepackage{multirow}
\usepackage{xcolor}
\usepackage{tcolorbox}
\usepackage{lipsum}
\usepackage{amsmath}
\usepackage{tikz}
\usepackage{subcaption}
\usepackage{algorithm}
\usepackage{algpseudocode}
\usepackage{subcaption}

\title{\includegraphics[width=0.15\linewidth]{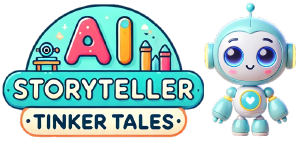} \texttt{Tinker Tales:} Interactive Storytelling Framework\\for Early Childhood Narrative Development and AI Literacy}

\author{
  \textbf{Nayoung Choi, $\;$ Peace Cyebukayire, $\;$ Jinho D. Choi} \\
  Department of Computer Science \\
  Emory University \\
  Atlanta, GA, USA \\
  \normalsize{\texttt{\{nayoung.choi, peace.cyebukayire, jinho.choi\}@emory.edu}}
}

\begin{document}
\maketitle
\begin{abstract}
This paper presents \textbf{\texttt{Tinker Tales}}, an interactive storytelling framework in the format of a board game, designed to support both narrative development and AI literacy in early childhood. The framework integrates tangible and speech-based interactions with AI through NFC chip-attached pawns and tokens, along with a speaker and microphone. Children select and define key story elements—such as characters, places, items, and emotions—using the pawns and tokens, providing further details to the AI and receiving proper assistance, similar to how adults prompt AI for specific tasks (e.g., writing). For evaluation, several game sessions were simulated with a child AI agent, and the quality and safety of the generated stories were assessed from various perspectives. This work highlights the potential of combining physical and digital elements in AI literacy, offering a safe and engaging way for children to learn how to effectively collaborate with AI.
\end{abstract}

\section{Introduction}

Artificial Intelligence (AI) technologies, represented by Large Language Models (LLMs), are being rapidly integrated into various domains, including education (\citealt{chen-etal-2024-dr}, \citealt{nayeem-rafiei-2024-kidlm}), healthcare (\citealt{zhu-etal-2025-integrating}, \citealt{fan-etal-2025-ai}), and entertainment (\citealt{fei-etal-2025-current}, \citealt{mahmud-marculescu-2024-opensep}). In the field of children's education, LLMs are increasingly being used to support various educational purposes such as problem-solving (\citealt{qi-etal-2025-tmath}) and literacy (\citealt{valentini-etal-2025-measuring}, \citealt{glandorf-meurers-2024-towards}). Recent studies in the Natural Language Processing (NLP) and Human-Computer Interaction (HCI) communities have highlighted the potential of LLMs in supporting children's narrative development\footnote{Narrative development refers to the stages in children's ability to understand and construct stories, which involves not only the structure of narratives but also the interpretation of the world around them \cite{applebee1978child}.} through interactive storytelling activities (\citealt{saraswat-etal-2024-story}, \citealt{10.1145/3613904.3642647}, \citealt{10.1145/3536221.3556578}), thereby fostering cognitive growth and creativity.

\begin{figure*}[t!]
    \centering
    \includegraphics[width=1.0\linewidth]{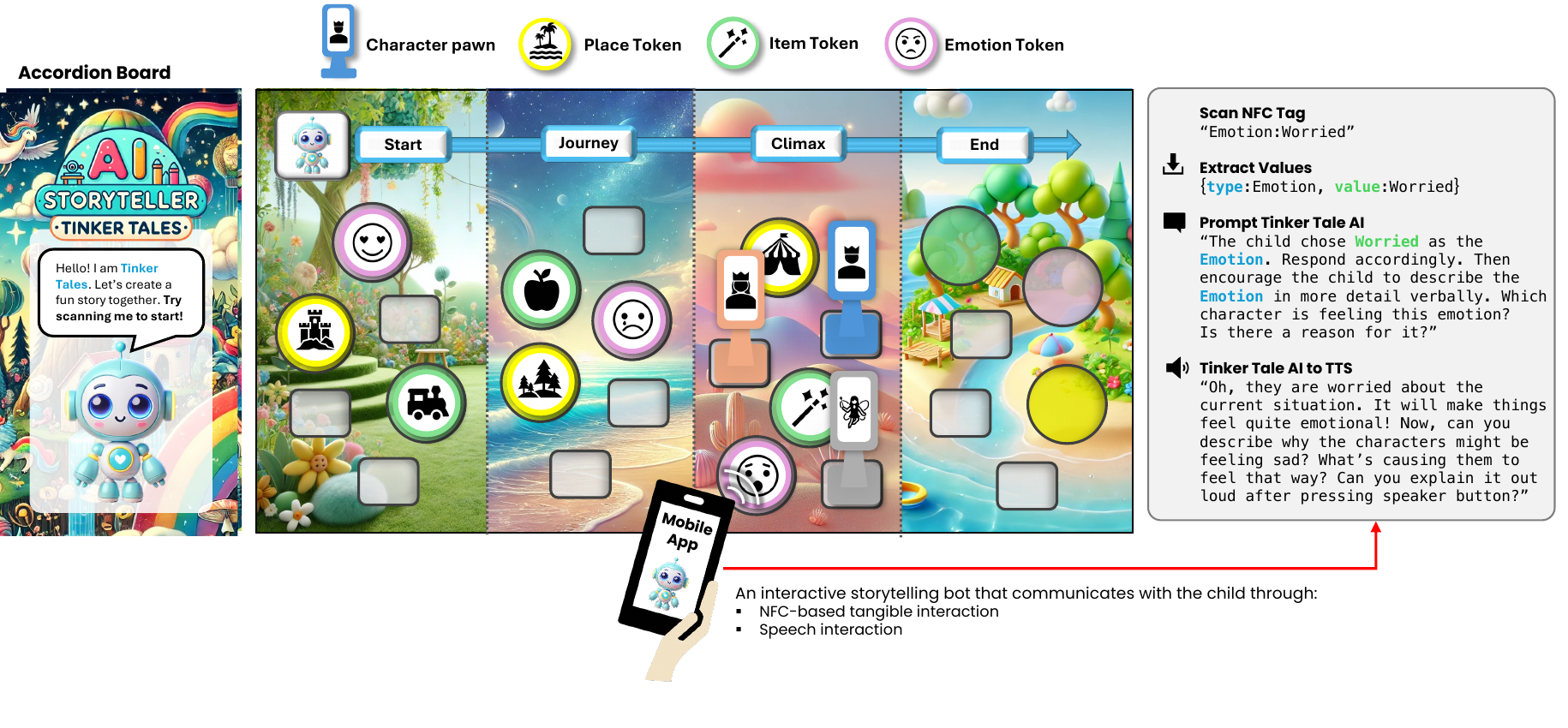}
    \vspace{-1cm}
    \caption{\textbf{\texttt{Tinker$\;$Tales}} features a four-sided accordion board, NFC chip-attached pawns and tokens, and a mobile app that mediates tangible and speech interaction. Guided by the AI, the user begins by selecting and defining \colorbox{blue}{\textcolor{white}{characters}}. Then, at each stage of the story, the user chooses and defines the \colorbox{yellow}{place}, \colorbox{green}{item}, and \colorbox{pink}{emotion} through interactions, shaping the narrative. Once the activity is completed, the entire story is saved, if desired by the user.}
    \label{fig:overview}
    \vspace{-0.5em}
\end{figure*}

Existing AI interactive storytelling activities are mostly confined to digital environments, often overlooking the benefits of physical interaction. Despite the growing reliance on screen-based learning, tactile engagement, which involves using multiple senses, remains crucial for cognitive development of younger children (\citealt{10.1145/3628516.3655751}, \citealt{10.1007/s00779-011-0404-2}, \citealt{10.1145/1810543.1810552}). Physical interaction with objects—such as toys, puzzles, or even simple gestures—plays an essential role in helping children grasp abstract concepts \cite{10.1145/1551788.1551839}.

Furthermore, despite the growing prevalence of AI-powered educational tools, there remains a gap in education aimed at helping children understand how AI systems work and how to effectively collaborate with them \cite{10.1145/3677613}. Adults learn to interact with AI through experiences where they prompt it for specific purposes. Children also need this type of experience in an AI-driven world. Through hands-on AI literacy education, especially with tasks like story generation where AI can help encourage creativity, children can learn to structure their thoughts and prompt the AI, while also receiving proper help to achieve specific goals.

This paper presents \textbf{\texttt{Tinker$\;$Tales}}, an interactive framework that provides children with an engaging storytelling experience aimed at narrative development, helping them learn how to collaborate with AI. Our system, as shown in Figure \ref{fig:overview}, consists of hardware (Section \ref{subsec:hardware}) and software (Section \ref{subsec:software}), moving beyond screen-based digital learning by supporting physical interaction through a game board with Near-Field Communication (NFC) \cite{10716601} chip-attached pawns and tokens. Children select and define characters and story elements, creating their own narratives while interacting with AI in both tangible and verbal ways. Our story generation and conversational agent approaches are described in Section \ref{sec:approach}. For the evaluation (Section \ref{sec:eval}), several game sessions were simulated using a child AI agent, and the quality and safety of the generated stories were assessed from various perspectives.


\section{Related Work}
\label{sec:related_work}

\textbf{Children and AI} $\;$ To improve backbone LLMs for child-centric applications, KidLM \cite{nayeem-rafiei-2024-kidlm} introduces a kid-appropriate text corpus and a novel masking process to enhance the model’s focus on child-specific words, addressing linguistic nuances and safety standards. \citet{bhandari-brennan-2023-trustworthiness} investigates LLM-generated content for children, using the OPT \cite{zhang2022optopenpretrainedtransformer}, LLaMA \cite{touvron2023llamaopenefficientfoundation}, and Alpaca \cite{alpaca} to generate children's stories and assess their safety (e.g., threats, insults, identity hate), thoroughly examining the appropriateness of LLMs in generating content for children. \citet{valentini-etal-2023-automatic} proposes LLMs prompting techniques to ensure that the vocabulary remains age-appropriate and safe for children.

\vspace{0.5em}
\noindent \textbf{Story Generation} $\;$ Story generation is fundamentally about defining what drives the narrative forward. \citet{brahman-etal-2020-cue} proposes a cue-driven story generation, where the narrative evolves based on specific cues provided by the user, which can range from a single word, enabling more controlled and dynamic storytelling. \citet{zhang-etal-2022-persona} focuses on persona-driven story generation, where the personality traits of characters shape the development of the story, leading to more personalized narratives. Additionally, \citet{vijjini-etal-2022-towards} explored interaction-driven story generation, emphasizing the importance of character interactions in guiding the direction of the plot. Event-driven story generation is another approach, with works like \citet{tang-etal-2022-etrica} and \citet{tang-etal-2022-ngep} examining how sequences of events can steer the narrative structure, ensuring coherence and logical progression in the generated stories.

\vspace{0.5em}
\noindent \textbf{Interactive Storytelling} $\;$ In both the NLP and HCI communities, interactive storytelling applications have gained attention. For example, Story-yarn \cite{saraswat-etal-2024-story} shows an interactive story building application that uses user-provided keywords, a story knowledge graph, and an LLM to dynamically create and select sentences. TIPES \cite{10.1145/3628516.3655751} explores how enactive actions and vocalizations help children construct story ideas through physical tabletop play. CreativeBot \cite{10.1145/3536221.3556578} introduces an AI bot that encourages children to contribute to the story, building it turn by turn, and fosters creativity. \citet{lee-etal-2022-interactive} presents a digital system that allows parents and children to interactively rewrite a story by answering questions to customize story details, with the option to accept or reject changes to the narrative.

\section{System Overview}
\label{sec:overview}
\begin{figure}[h!]
    \centering
    \includegraphics[width=1.0\linewidth]{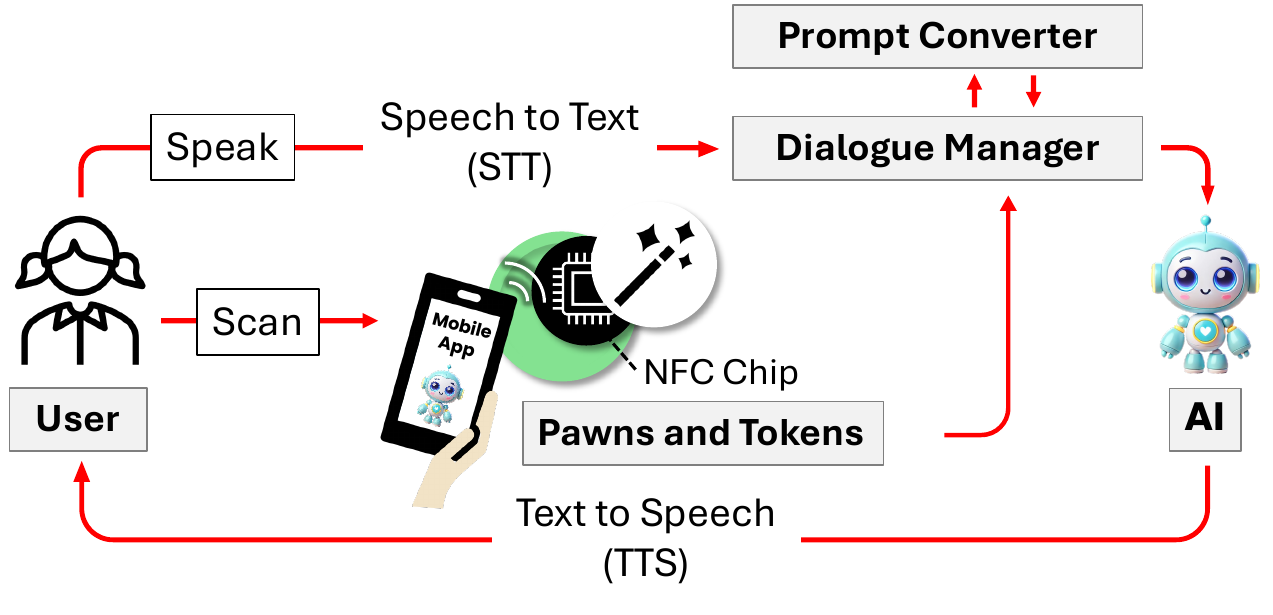}
    \caption{The user scans NFC chip-attached pawns and tokens or provides speech input. The NFC data or speech input, after STT processing, is converted into a proper prompt by the Dialogue Manager and sent to the AI. The AI generates an suitable response or instruction, which is then delivered to the user through TTS.}
    \label{fig:components}
\vspace{-1em}
\end{figure}

Our system consists of hardware (Section \ref{subsec:hardware}) and software (Section \ref{subsec:software}) components, with the user interaction flow shown in Figure \ref{fig:components}.

\subsection{Hardware}
\label{subsec:hardware}
\begin{figure}[h!]
    \centering
    \includegraphics[width=1.0\linewidth]{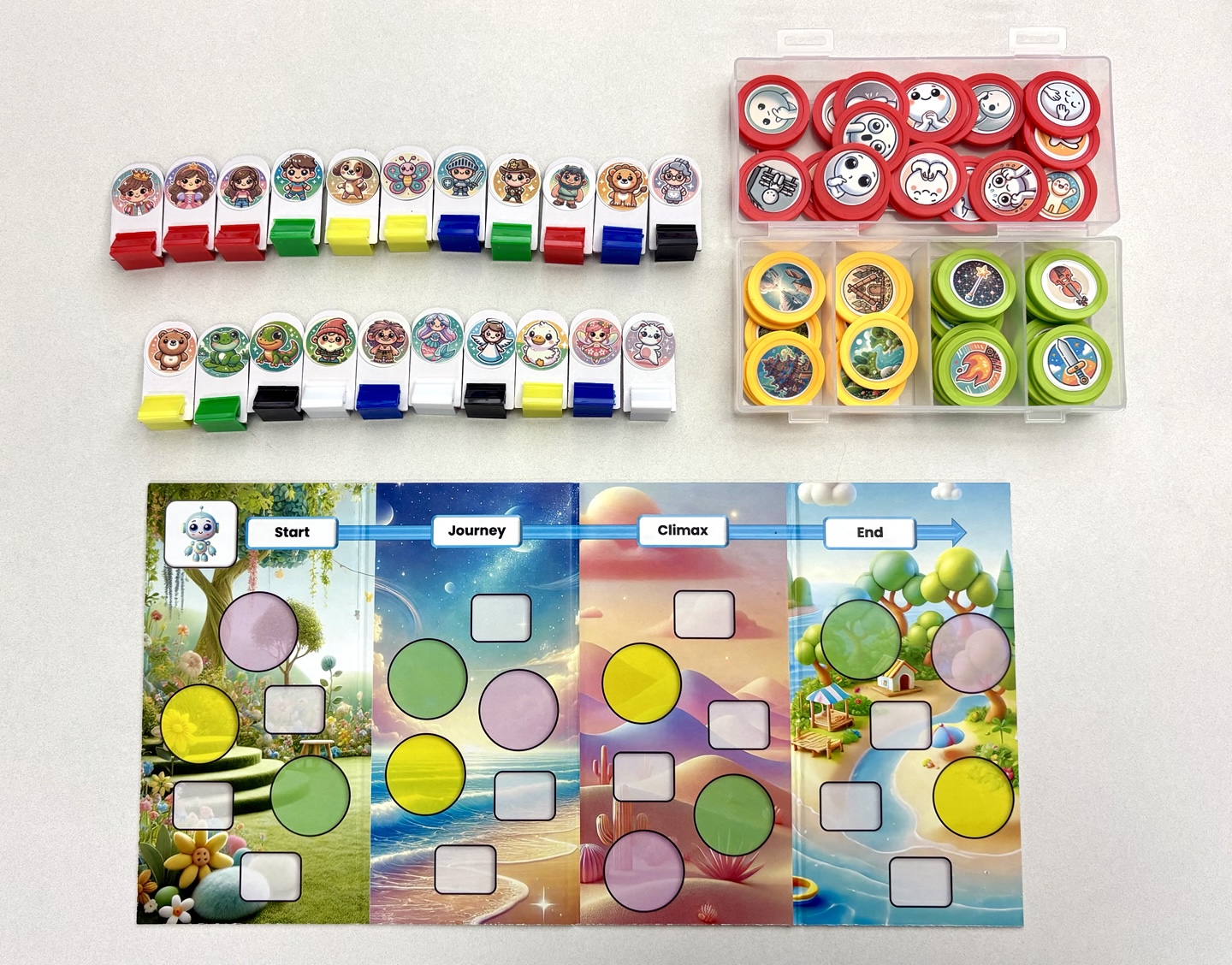}
    \caption{The physical setup of our system, showing the game board and associated pawns and tokens.}
    \label{fig:photo}
\vspace{-1em}
\end{figure}

The hardware includes a customized game board, pawns and tokens, and a mobile phone for the NFC chip scanning, speaker and microphone. A photo of the physical setup is shown in Figure \ref{fig:photo}.

\vspace{0.5em}
\noindent \textbf{Game Board} $\;$ The game board is designed as a four-sided accordion board, divided into Start, Journey, Climax and End pages. The user moves sequentially through the pages, placing pawns and tokens as they progress, completing the story in order through the activity. The images used in the board design were all generated by DALL·E 3\footnote{\url{https://openai.com/index/dall-e-3/}}, and the finished custom design was made to order\footnote{\url{https://www.thegamecrafter.com/make/products/AccordionBoard}}.

\vspace{0.5em}
\noindent \textbf{Pawns and Tokens} $\;$ There are one type of pawn (character) and three types of tokens (place, item and emotion). The selection options for character, place, item, and emotion were extracted from the Children Stories Text Corpus\footnote{\url{https://www.kaggle.com/datasets/edenbd/children-stories-text-corpus}} using GPT-4o\footnote{\url{https://platform.openai.com/docs/models/gpt-4o}}. The extracted list is in Table \ref{tab:selection}, with further details on the extraction process available in \ref{subsec:selection}. Each pawn and token has a NFC chip\footnote{We used the NTAG215 type of NFC chip, which has 540 bytes of user memory and offers read and write capabilities.} attached. The user scans the pawns and tokens with their mobile phone running our app, which has NFC capabilities, and then places them on the game board. The pawns, tokens, NFC chips, and printable sticker papers were ordered from Amazon\footnote{\url{https://www.amazon.com/}} and handmade. All sticker images were generated by DALL·E 3.

\begin{table}[h!]
\renewcommand{\arraystretch}{1.5} 
\centering
\small
\begin{tabular}{|p{0.2\linewidth}|p{0.7\linewidth}|} 
\hline
\multicolumn{1}{|c|}{\textbf{Type}} & \multicolumn{1}{c|}{\textbf{Options}} \\ \hline
\centering \texttt{Character} & Prince, Princess, Girl, Boy, Nanny, Knight, Sheriff, Giant, Troll, Dwarf, Mermaid, Fairy, Angel, Lion, Butterfly, Duck, Frog, Lizard, Dog, Bear, Rabbit \\ \hline
\centering \texttt{Place} & Forest, Island, Cave, Castle, Garden, River, Street, Theatre, Bridge, Harbour, Hut, Market, Mountain, Temple  \\ \hline
\centering \texttt{Item} & Wand, Ruby, Matches, Sword, Flower, Fire, Ring, Lantern, Coins, Boots, Cake, Violin, Harp, Thorn, Book, Ladder, Stick, Clock, Apple, Doll, Hat, Shoes, Ship, Crown \\ \hline
\centering \texttt{Emotion} & Happy, Sad, Lonely, Loving, Joyful, Painful, Passionate, Regretful, Angry, Proud, Anxious, Disappointed, Suspicious, Satisfied, Embarrassed, Hopeful, Fearful, Curious, Comfortable, Reflective, Surprised, Scared, Thankful \\ \hline
\end{tabular}
\caption{The list of available options for the character, place, item, and emotion (\ref{subsec:selection}). The user's choices and definitions  drive the story generation in our system.}
\vspace{-1em}
\label{tab:selection}
\end{table}


\vspace{0.5em}
\noindent \textbf{Mobile Phone} $\;$ The mobile phone runs our software app (Section \ref{subsec:software}) and serves as the device for the NFC chip scanning, speaker and microphone.

\subsection{Software}
\label{subsec:software}

\begin{figure}[h!]
    \centering
    \includegraphics[width=1.0\linewidth]{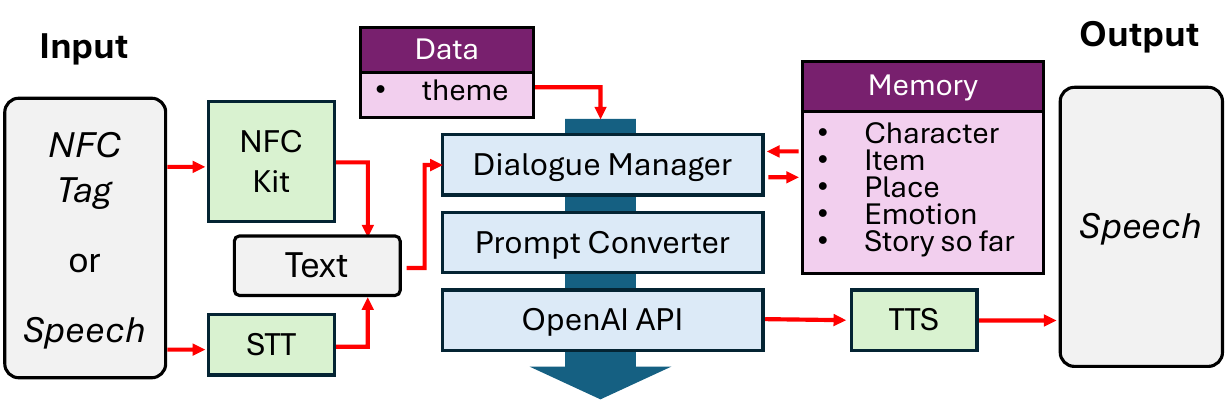}
    \caption{Our mobile app provides appropriate feedback and instructions through speech, based on the NFC chip or speech input provided by the user. The progression of the game is managed by the Dialogue Manager (\ref{subsec:dialogue_manager}).}
    \label{fig:software}
\vspace{-1em}
\end{figure}

Figure \ref{fig:software} illustrates the architecture of our mobile app, developed using Flutter, which is deployable on both iPhone and Android. When the user provides NFC chip or speech input, it is converted into text using the NFC Kit\footnote{\url{https://pub.dev/packages/flutter\_nfc\_kit}} or STT\footnote{\url{https://pub.dev/packages/speech\_to\_text}}, respectively. The converted text is then passed to Dialogue Manager (\ref{subsec:dialogue_manager}). The Dialogue Manager controls the game progression using a pointer, which keeps track of the current stage in the game. It also stores the user’s choices (e.g., characters, items, places and emotions) and the story created so far, referencing them as needed. The Prompt Converter transforms the input text into the proper prompt based on the pointer’s position. The prompt is sent as a request to the OpenAI API\footnote{\url{https://openai.com/index/openai-api}}, and the response, which is the proper instructions for progressing the game, is delivered to the user through TTS\footnote{\url{https://pub.dev/packages/flutter\_tts}}. This process repeats until the story is complete. Additionally, our app features a function allowing parents to pre-enter one-line lessons (e.g., "Do not lie," "Get along with friends"), with one of these lessons being selected as the theme for the story generation.

\vspace{0.5em}
\noindent \checkmark $\;$ Since our system requires custom hardware, including a game board, pawns and tokens, we are unable to provide a live website. However, a demonstration video\footnote{\url{https://www.youtube.com/watch?v=DdfhTg7QlVU}} and an additional uncut version video\footnote{\url{https://www.youtube.com/watch?v=zW54TSVsEFE}} fully showcase the system in action.

\section{Apporach}
\label{sec:approach}

\noindent \textbf{Story Generation} $\;$ The characteristic of our story generation approach is that the user selects and defines all story elements (see Table \ref{tab:selection_example}), and the AI only provides logical and creative connections between them. To ensure the story remains coherent, even with random user choices, our framework is based on a 4-stage story structure (introduction, development, crisis, conclusion). The AI model is prompted with the current stage of the story, allowing it to generate content that maintains coherence with the user's choices within that stage. (e.g., \textit{"Make sure the place, item, and emotion chosen by the child are naturally incorporated into the plot. $\ldots$ , The next part can take place in a completely different location, so the events occurring in this part should be concluded here."}). Additionally, as detailed in Table \ref{tab:prompt} (\ref{subsec:dialogue_manager}), prompt engineering was conducted to ensure smooth transitions between stages of the story (e.g., \textit{"You need to explain the transitions between places naturally, ..., in a way that feels seamless and cohesive."}).

\vspace{0.5em}
\noindent \textbf{Conversational Agent} $\;$ Our conversational agent, \texttt{\textbf{Tinker Tale}} AI, operates by following pre-defined pointers from the Dialogue Manager, such as \texttt{Initiate} > \texttt{Character:1:Select} > \texttt{Character:1:Define}, $\ldots$ , and so on. The Dialogue Manager takes the current user input (NFC chip or speech) and Prompt Converter constructs a prompt to instruct the AI (e.g., \textit{"The child chose \texttt{Prince} as the first character. Ask the child to verbally say the name they want to give and describe the character's traits."}). Additionally, to ensure child-appropriate interaction, we use system prompts (e.g., \textit{"You are a friendly intelligent agent named Tinker Tales, helping a child aged 4-6 with story generation using a game board."}) and provide reminders throughout (e.g., \textit{"Use simple words and sentence structures, keeping in mind that the listener is a child aged 4-6."}).

\section{Evaluation}
\label{sec:eval}
\begin{figure}[h!]
    \centering
    \includegraphics[width=1.0\linewidth]{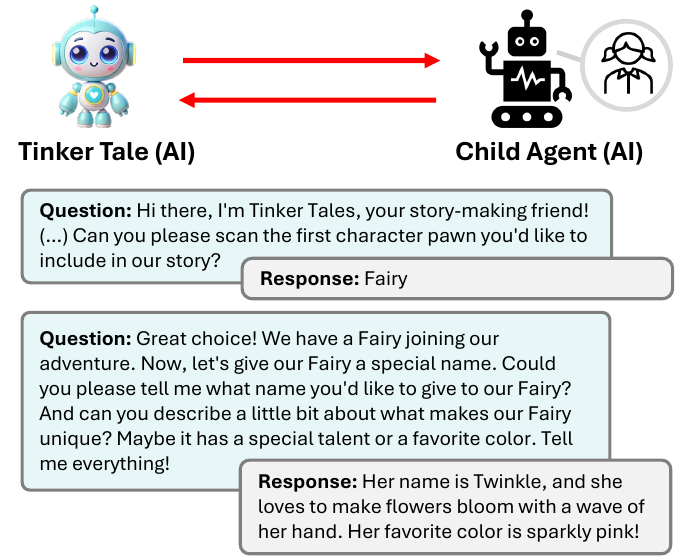}
    \caption{Example interaction between the \texttt{\textbf{Tinker Tale}} AI and the child AI agent, demonstrating the process of selecting and defining the first character.}
    \label{fig:eval}
\vspace{-1em}
\end{figure}

To evaluate the generated stories in various settings by our system, we simulated 30 game sessions with an AI agent role-playing as a child user, as instructed in Figure \ref{fig:prompt_child_agent} (\ref{subsec:evaluation_questionnaire}), using GPT-4o. The AI child agent, mimicking a child, randomly selects and defines a story element each turn by responding to our \texttt{\textbf{Tinker Tale}} AI, as shown in Figure \ref{fig:eval}. For example, Table \ref{tab:selection_example} illustrates a simulation where the AI child agent makes selections and definitions during a session, while Table \ref{tab:story_example} presents the full story generated by our system from that session.

\begin{table*}[t!]
\renewcommand{\arraystretch}{1.5} 
\centering
\scriptsize
\begin{tabular}{|p{0.08\linewidth}|p{0.06\linewidth}|p{0.06\linewidth}|p{0.7\linewidth}|} 
\hline
\multicolumn{1}{|c|}{\textbf{Stage}} & \multicolumn{1}{|c|}{\textbf{Element}} & \multicolumn{1}{|c|}{\textbf{Selection}} & \multicolumn{1}{c|}{\textbf{Selection Defined by Child AI Agent}}                                                                                                                                                                                                                                                                                                                                                   \\ \hline
\centering \multirow{3}{*}{\textcolor{gray}{\texttt{Character}}}    & First         & Troll              & His name is Grumble, and he can make flowers grow with his footsteps!                                                                                                                                                                                                                                                                                                                           \\ \cline{2-4} 
                              & Second        & Lion               & His name is Leo, and he can talk to the wind!                                                                                                                                                                                                                                                                                                                                                   \\ \cline{2-4} 
                              & Third         & Mermaid            & Her name is Pearl, and she can turn raindrops into tiny fish!                                                                                                                                                                                                                                                                                                                                   \\ \hline
\centering \multirow{3}{*}{\texttt{\textcolor{orange}{Start}}} & Place         & Bridge             & The bridge is made of twisting vines and colorful flowers. It glows with little sparkly lights at night, and there's a gentle stream flowing beneath it that whispers when the wind blows. It feels peaceful and magical!                                                                                                                                                                       \\ \cline{2-4} 
                              & Item          & Hat                & The hat is a big, floppy one with colorful patches and a feather on top. Grumble found it under the bridge, and when he puts it on, it makes everyone laugh by tickling them with magic!                                                                                                                                                                                                        \\ \cline{2-4} 
                              & Emotion       & Sad                & Pearl is feeling sad because she misses her home in the ocean.                                                                                                                                                                                                                                                                                                                                  \\ \hline
\centering \multirow{3}{*}{\texttt{\textcolor{teal}{Journey}}}  & Place         & Street             & The street is busy and full of colorful market stalls with lots of people. There are happy sounds everywhere, like music and people talking, and it smells like yummy food. It feels lively and exciting!                                        \\ \cline{2-4} 
                              & Item          & Cake               & They found the cake at a market stall. It's big and has rainbow sprinkles on top, and when you eat it, it makes you giggle and feel really happy! \\ \cline{2-4} 
                              & Emotion       & Anxiety            & Leo is feeling anxious because he's not used to so many people and noise around.                                                                                                                                                                                                                                                                                                                \\ \hline
\centering \multirow{3}{*}{\texttt{\textcolor{violet}{Climax}}}       & Place         & Mountain           & The mountain is tall and covered with snow at the top. There are lots of trees that look like soldiers standing in a row, and the air is fresh and crisp. It feels adventurous and a bit mysterious!                                                                                                                                                                                            \\ \cline{2-4} 
                              & Item          & Apple              & They found the apple on a tree growing near the top of the mountain, and it's shiny and golden with little sparkles. When you eat it, it gives you courage!                                                                                                                                                                                                                                      \\ \cline{2-4} 
                              & Emotion       & Happy              & Grumble is feeling happy because he loves being in nature and loves the fresh mountain air. It makes him dance!                                                                                                                                                                                                                                                                                 \\ \hline
\centering \multirow{3}{*} {\texttt{\textcolor{magenta}{End}}}   & Place         & Island             & The island is small and surrounded by clear blue waters. There's a sandy beach with shells that sparkle in the sun, and palm trees that sway gently in the breeze. It feels calm and magical, like a secret paradise!                                                                                                                                                                           \\ \cline{2-4} 
                              & Item          & Boots              & They found the boots washed up on the shore. They're made of golden fabric and, when worn, make you feel like you can jump really high!                                                                                                                                                                                                                                                         \\ \cline{2-4} 
                              & Emotion       & Fear               & Pearl is feeling a bit afraid because she misses being surrounded by water, and the island is different from her ocean home.                                                                                                                                                                                                                                                                    \\ \hline
\end{tabular}
\caption{The story elements selected and defined by the child AI agent during a simulated game session.}
\vspace{-0.5em}
\label{tab:selection_example}
\end{table*}

\begin{table*}[t!]
\renewcommand{\arraystretch}{1.5} 
\centering
\scriptsize
\begin{tabular}{|p{0.078\linewidth}|p{0.88\linewidth}|} 
\hline
\multicolumn{1}{|c|}{\textbf{Stage}} & \multicolumn{1}{c|}{\textbf{Generated Story}} \\ \hline
\centering \texttt{\textcolor{orange}{Start}} & Once upon a time, at a magical bridge made of twisting vines and colorful flowers, Grumble the Troll stepped carefully, leaving a trail of blooming blossoms behind him. He was joined by his friend Leo, the Lion, who could speak to the wind, and Pearl, the Mermaid, who was a little sad because she missed her home in the ocean. As the three friends chatted near the gentle stream whispering beneath the bridge, Grumble happened upon a big, floppy hat covered in colorful patches, with a feather perched on top. He picked it up and, with a smile, placed it on his head. To everyone's surprise, magical tickles spread through the air, causing them all to burst into laughter. Even Pearl, who had been feeling blue, found herself giggling at the silly, tickly magic. Seeing Pearl smile, Grumble and Leo decided to use the hat's magic to cheer her up whenever she felt sad. The three friends grew closer as they spent the day at the enchanting bridge, surrounded by sparkling lights and laughter, helping Pearl feel a little more at home in this wondrous place. \\ \hline
\centering \texttt{\textcolor{teal}{Journey}} & As the sun began to set, Grumble, Leo, and Pearl decided to explore the bustling street nearby. The street was full of colorful market stalls and lively sounds of music and laughter. The air was filled with the delicious smell of tasty treats. While Grumble and Pearl were excited by the lively atmosphere, Leo felt a little anxious with all the people and noise around. But just then, they discovered a special market stall selling a magical cake sprinkled with rainbow goodness. Grumble bought the cake, hoping it would cheer Leo up. When Leo took a bite of the cake, he began to giggle uncontrollably, and his anxious feelings started to melt away as happiness bubbled up inside him. His laughter was contagious, spreading to Grumble and Pearl, turning Leo's nervousness into joy. Surrounded by friendly faces, yummy food, and the delightful cake, Leo realized he was safe and happy with his friends by his side. In this lively place, their friendship grew even stronger, and Leo learned that with the support of friends, he could embrace new experiences with joy. \\ \hline
\centering \texttt{\textcolor{violet}{Climax}} & On a bright and crisp morning, Grumble, Leo, and Pearl climbed the snowy mountain, where the air was fresh and full of adventure. Grumble, feeling happy to be surrounded by nature, twirled around in a joyful dance. As they explored, they found a shining golden apple sparkling on a tree near the top. Curious and eager, Grumble plucked the apple and shared it with his friends. As they each took a bite, a warm feeling of courage filled their hearts. Just then, a thick fog rolled in, making it hard to see the path back down the mountain. Grumble, Leo, and Pearl huddled together, uncertain of where to go. But they remembered the courage from the magical apple and held hands, feeling brave. With Grumble's happiness, Leo's ability to talk to the wind, and Pearl's gentle voice calming their fears, they decided to move forward together through the fog. Leo whispered to the wind, asking it for guidance, and they slowly started down the mountain. The wind, as a helpful friend, listened and created a path through the fog. Together, the friends found their way back to safety with joyful hearts, knowing they had each other in times of trouble. And on that mountain, under the clear sky, their friendship shone brighter than ever. \\ \hline
\centering \texttt{\textcolor{magenta}{End}} & Arriving on the small, magical island surrounded by clear blue waters, Grumble, Leo, and Pearl felt a sense of calm wash over them. The sandy beach sparkled with seashells that glistened in the sun, and the palm trees gently swayed in the warm breeze. It truly felt like a secret paradise. While exploring the beach, they stumbled upon a pair of boots washed up on the shore. The boots were made of golden fabric, glistening as brilliantly as the shells. When Grumble tried them on, he found he could jump higher than ever before, and his joy turned into delighted laughter. Pearl watched Grumble jumping happily, but she felt a bit of fear inside. She missed being surrounded by water and the feeling of home in the ocean. With a kind smile, Leo used his voice to speak to the gentle waves, and the ocean sang a soft melody to comfort Pearl. Grumble, realizing Pearl’s worry, quickly shared the magic boots with her. As she wore them, she felt a lift not only in her jump but in her spirit, surrounded by the understanding and support of her friends. Together, they played on the sparkling beach, their friendship wrapping them in warmth and joy. And so, on this magical island, Grumble, Leo, and Pearl discovered that true happiness was found in sharing and supporting one another, appreciating the present moment, no matter where they were. \\ \hline
\end{tabular}
\caption{The generated full story from the same simulated game session in Figure \ref{tab:selection_example}.}
\vspace{-1.5em}
\label{tab:story_example}
\end{table*}

\noindent The important factors in evaluating the stories generated by our system are that the user's element selections and definitions should be seamlessly integrated into each stage of the story, and the main narrative should flow logically from start to end. Additionally, the lessons provided by the parent should be reflected as the theme throughout the story. Finally, the content should be free from any harm to the child. The four evaluation criteria we defined are as follows:
\vspace{-0.2em} 
\begin{itemize} 
\setlength\itemsep{-0.1em} 
\item \textbf{Elements Relevancy} measures whether the elements defined by the user are included in the story's plot and play an important role.
\item \textbf{Narrative Coherence} evaluates whether the story follows a logical progression from introduction to conclusion, maintaining consistency and a clear flow of events throughout. 
\item \textbf{Educational Value} assesses whether the story effectively conveys the intended lessons and encourages positive learning outcomes.
\item \textbf{Safety} examines whether the story is free from harmful, offensive, or inappropriate language, ensuring that it is suitable for children.
\end{itemize}

\noindent For quality evaluation, which includes elements relevancy, narrative coherence, and educational value, we create an evaluation questionnaire (Table \ref{fig:questionnaire} in \ref{subsec:evaluation_questionnaire}), where each aspect is answered with Yes or No and rated on a 1-5 scale. Using this questionnaire, we conduct both human evaluation and LLM-as-a-judge evaluation \cite{zheng2023judgingllmasajudgemtbenchchatbot}. For safety evaluation, we use the Moderation API\footnote{\url{https://moderationapi.com}} and Perspective API\footnote{\url{https://perspectiveapi.com}} to analyze whether our stories contain harmful or inappropriate content.

\subsection{Quality Evaluation}
\label{subsec:llm_eval}
Table \ref{tab:eval_quality_llm} shows the results of human and LLM evaluations for the questionnaire (Figure \ref{fig:questionnaire}). Our system generates coherent and educational stories in line with the user’s instructions, with all metrics averaging above 4.5 on a 5-point scale in both evaluations.

\begin{table}[h!]
\centering
\small
\centering
\resizebox{0.7\columnwidth}{!}{%
\begin{tabular}{crr}
\toprule
\textbf{Metric}     & \multicolumn{1}{c}{\textbf{Human}} & \multicolumn{1}{c}{\textbf{LLM}} \\ \midrule
Elements Relevancy  & \begin{tabular}[c]{@{}r@{}}5.00\\ \scriptsize{($\pm$ 0.0)}\end{tabular} & \begin{tabular}[c]{@{}r@{}}4.53\\ \scriptsize{($\pm$ 0.50)}\end{tabular} \\
Narrative Coherence & \begin{tabular}[c]{@{}r@{}}4.77\\ \scriptsize{($\pm$ 0.43)}\end{tabular} & \begin{tabular}[c]{@{}r@{}}4.80\\ \scriptsize{($\pm$ 0.40)}\end{tabular} \\
Educational Value   & \begin{tabular}[c]{@{}r@{}}4.97\\ \scriptsize{($\pm$ 0.18)}\end{tabular} & \begin{tabular}[c]{@{}r@{}}4.53\\ \scriptsize{($\pm$ 0.50)}\end{tabular} \\ \bottomrule
\end{tabular}%
}
\vspace{0.1em}
\caption{Human and LLM Evaluation Results for the questionnaire assessing three quality metrics (1-5 scale).}
\vspace{-0.5em}
\label{tab:eval_quality_llm}
\end{table}

\noindent Our story generation approach (Section \ref{sec:approach}), which ensures coherence within each of the 4-stage story structure and strengthens the logical connections between stages, allows the narrative to maintain coherence even with random user choices. For instance, a sudden place change from a mountain to a theatre is smoothly handled. The introduction starts with: \textit{"Once upon a time, high up on a tall and snowy mountain, $\ldots$ , they discovered a shiny, magical ring. $\ldots$ , 'This must lead to something special,' Pearl thought aloud."} The development follows with: \textit{"Following the path that the magical ring hinted at, they left the snowy mountain behind and arrived at a grand and lively theatre."}

\subsection{Safety Evaluation}
\label{subsec:safety_eval}
Table \ref{tab:eval_safety} compares the content safety of stories generated by our system with 30 samples from the Children Stories Text Corpus (used for story element extraction in Section \ref{subsec:hardware}), evaluated using the Moderation API and Perspective API. Both sources exhibited very low harmfulness across all safety categories (scored on a 0-1 scale), with our stories showing lower harmfulness in every category.

\begin{table}[h!]
\centering
\small
\begin{subtable}[b]{0.45\textwidth}
\centering
\resizebox{0.8\columnwidth}{!}{%
\begin{tabular}{c|rr}
\toprule
\textbf{}       & \multicolumn{1}{c}{\textbf{\texttt{Tinker Tale}}}                                & \multicolumn{1}{c}{\textbf{\begin{tabular}[c]{@{}c@{}}Children Stories \\ Text Corpus\end{tabular}}} \\ \midrule
Harassment        & \begin{tabular}[c]{@{}r@{}}0.0010\\ \scriptsize{($\pm$ 0.0015)}\end{tabular}  & \begin{tabular}[c]{@{}r@{}}0.0272\\ \scriptsize{($\pm$ 0.0458)}\end{tabular}                                      \\
Hate & \begin{tabular}[c]{@{}r@{}}0.0\\ \scriptsize{($\pm$ 0.0)}\end{tabular}  & \begin{tabular}[c]{@{}r@{}}0.0023\\ \scriptsize{($\pm$ 0.0032)}\end{tabular}                                      \\
Illicit & \begin{tabular}[c]{@{}r@{}}0.0\\ \scriptsize{($\pm$ 0.0)}\end{tabular}  & \begin{tabular}[c]{@{}r@{}}0.0027\\ \scriptsize{($\pm$ 0.0034)}\end{tabular}                                      \\
Self Harm       & \begin{tabular}[c]{@{}r@{}}(0.0038\\ \scriptsize{($\pm$ 0.0044)}\end{tabular} & \begin{tabular}[c]{@{}r@{}}0.0254\\ \scriptsize{($\pm$ 0.0758)}\end{tabular}                                      \\
Sexual          & \begin{tabular}[c]{@{}r@{}}0.0064\\ \scriptsize{($\pm$ 0.0072)}\end{tabular}  & \begin{tabular}[c]{@{}r@{}}0.0201\\ \scriptsize{($\pm$ 0.0277)}\end{tabular}                                      \\
Violence          & \begin{tabular}[c]{@{}r@{}}0.0519\\ \scriptsize{($\pm$ 0.0765)}\end{tabular}  & \begin{tabular}[c]{@{}r@{}}0.1647\\ \scriptsize{($\pm$ 0.2246)}\end{tabular}                                      \\ \bottomrule
\end{tabular}%
}
\vspace{-0.1cm}
\caption{Moderation API (Avg.)}
\end{subtable}
\vspace{0.3cm}

\begin{subtable}[b]{0.45\textwidth}
\centering
\resizebox{0.8\columnwidth}{!}{%
\begin{tabular}{c|rr}
\toprule
\textbf{}       & \multicolumn{1}{c}{\textbf{\texttt{Tinker Tale}}}                                & \multicolumn{1}{c}{\textbf{\begin{tabular}[c]{@{}c@{}}Children Stories \\ Text Corpus\end{tabular}}} \\ \midrule
Toxicity        & \begin{tabular}[c]{@{}r@{}}0.0695\\ \scriptsize{($\pm$ 0.0399)}\end{tabular}  & \begin{tabular}[c]{@{}r@{}}0.1301\\ \scriptsize{($\pm$ 0.0796)}\end{tabular}                                      \\
Identity Attack & \begin{tabular}[c]{@{}r@{}}0.0070\\ \scriptsize{($\pm$ 0.0047)}\end{tabular}  & \begin{tabular}[c]{@{}r@{}}0.0294\\ \scriptsize{($\pm$ 0.0445)}\end{tabular}                                      \\
Severe Toxicity & \begin{tabular}[c]{@{}r@{}}0.0026\\ \scriptsize{($\pm$ 0.0011)}\end{tabular}  & \begin{tabular}[c]{@{}r@{}}0.0061\\ \scriptsize{($\pm$ 0.0047)}\end{tabular}                                      \\
Profanity       & \begin{tabular}[c]{@{}r@{}}(0.0301\\ \scriptsize{($\pm$ 0.0149)}\end{tabular} & \begin{tabular}[c]{@{}r@{}}0.0688\\ \scriptsize{($\pm$ 0.0594)}\end{tabular}                                      \\
Threat          & \begin{tabular}[c]{@{}r@{}}0.0099\\ \scriptsize{($\pm$ 0.0023)}\end{tabular}  & \begin{tabular}[c]{@{}r@{}}0.0298\\ \scriptsize{($\pm$ 0.0581)}\end{tabular}                                      \\
Insult          & \begin{tabular}[c]{@{}r@{}}0.0331\\ \scriptsize{($\pm$ 0.0278)}\end{tabular}  & \begin{tabular}[c]{@{}r@{}}0.0621\\ \scriptsize{($\pm$ 0.0834)}\end{tabular}                                      \\ \bottomrule
\end{tabular}%
}
\vspace{-0.1cm}
\caption{Perspective API (Avg.)}
\end{subtable}
\vspace{0.1cm}
\caption{Comparison of content safety evaluation between \textbf{\texttt{Tinker Tale}} and Children Stories Text Corpus.}
\vspace{-0.5cm}
\label{tab:eval_safety}
\end{table}

\section{Conclusion}
This paper presented an interactive storytelling framework designed to support early childhood narrative development and AI literacy. Our system effectively facilitates collaboration with AI through speech and tangible interaction. Moreover, the stories generated through these activities demonstrated high quality and safety, making the system both engaging and educational for young users. Future work will focus on expanding the system to support reading comprehension activities based on the generated stories, offering children a more comprehensive educational experience that integrates storytelling with text understanding.

\section*{Acknowledgments}
We would like to thank Dr. Williams for introducing The IoT Codex~\cite{10305681}, a lightweight framework for integrating radio-frequency identification (RFID) with everyday physical objects, which inspired the early development of this work.

\bibliography{acl_latex}

\clearpage
\appendix
\section{Appendix}
\subsection{Characters, Places, Items and Emotions}
\label{subsec:selection}
As briefly mentioned in Section \ref{subsec:hardware}, the characters, places, items, and emotions used as options in the game were all extracted from the \textit{Children Stories Text Corpus}, a manually curated dataset consisting of public domain children's books from Project Gutenberg\footnote{\url{https://www.gutenberg.org/}}. We extracted the options from the first 30 stories of the dataset in order, and the titles of the stories are as follows. The prompt used for this process is shown in Figure \ref{fig:prompt_extract}.

\vspace{0.5em}
\noindent \textbf{Stories Used} $\;$ \textit{The happy prince; The nightingale and the rose; The selfish giant; The devoted friend; The remarkable rocket; The emperor's new clothes; The swineherd; The real princess; The shoes of fortune; The fir tree; The leap-frog; The bell; The old house; The happy family; The story of a mother; The false collar; The little match girl; The naughty boy; The bronze ring; Prince hyacinth and the dear little princess; The yellow dwarf; Little red riding hood; Cinderella; Aladdin and the wonderful lamp; Beauty and the beast; The master-maid; Why the sea is salt; Felicia and the pot of pinks; The white cat; The water-lily the gold-spinners; The terrible head; The story of pretty goldilocks}.

\begin{figure}[h!]
    \small
    \begin{tcolorbox}[width=\linewidth, colback=white, colframe=black, title=\textsc{Prompt}]
    
    Given a children's story, please extract as many characters, places, items and emotions as possible from the story. Characters, places, and items should be general terms consisting of a single word and should be simple and easy for children to understand. Also, summarize the lesson of the story in one short sentence. \\
    
    Story = \textcolor{blue}{\texttt{\{\{STORY\}\}}}
    \end{tcolorbox}
\caption{The prompt used to extract characters, places, items, and emotions in simple terms from a story.}
\vspace{-1em}
\label{fig:prompt_extract}
\end{figure}

\subsection{Evaluation}
\label{subsec:evaluation_questionnaire}
Figure \ref{fig:prompt_child_agent} is the system prompt used to instruct the AI agent role-playing as a child user during the simulation of game sessions. Figure \ref{fig:questionnaire} presents a questionnaire for the quality evaluation of the stories generated by our system (Section \ref{sec:eval}).

\begin{figure}[h!]
    \small
    \begin{tcolorbox}[width=\linewidth, colback=white, colframe=black, title=\textsc{System Prompt}]
    
    You are a role-player simulating a child aged 4-6 playing an interactive storytelling board game. You will follow the AI's guidance throughout the game, taking actions such as scanning a pawn or token or responding verbally on each turn. When it's time to scan, simply return the selected option without adding anything extra (e.g., Queen), and when responding verbally, just answer in natural conversational language. \\
    
    The items that can be scanned are as follows:
    
    \hspace{0.5em}Characters = \textcolor{blue}{\texttt{\{\{CHARACTERS\}\}}}
    
    \hspace{0.5em}Places = \textcolor{blue}{\texttt{\{\{PLACES\}\}}}
    
    \hspace{0.5em}Items = \textcolor{blue}{\texttt{\{\{ITEMS\}\}}}
    
    \hspace{0.5em}Emotions = \textcolor{blue}{\texttt{\{\{EMOTIONS\}\}}}

    \end{tcolorbox}
\caption{The system prompt used to instruct an AI agent role-playing as a child user in the game.}
\vspace{-0.5em}
\label{fig:prompt_child_agent}
\end{figure}

\begin{figure}[h!]
    \small
    \begin{tcolorbox}[width=\linewidth, colback=white, colframe=black, title=\textsc{Questionnaire}]
    \textbf{(1) Are all the elements defined by the user (characters, places, items, and moods) included in the story's plot?} \\
    
    \textsc{Yes / No} \\
    
    If yes, to what degree are they important to the plot? \\
    
    \texttt{[Scale]} 1: Not Important at All, 2: Slightly Important, 3:Moderately Important, 4:Very Important, 5:Crucial to the Plot \\

    \textbf{(2) Does the story maintain a logical progression from introduction to conclusion?} \\

    \textsc{Yes / No} \\

    If yes, how clear is the story’s flow of events? \\
    
    \texttt{[Scale]} 1: Very Confusing, 2: Somewhat Confusing, 3: Neutral, 4: Clear, 5: Very Clear \\
    
    \textbf{(3) Does the story clearly reflect one of the pre-defined lessons as its central theme?} \\
    
    \textsc{Yes / No} \\
    
    If yes, to what degree is the lesson effectively conveyed through the story? \\
    
    \texttt{[Scale]} 1: Not Conveyed at All, 2: Slightly Conveyed, 3: Moderately Conveyed, 4: Well Conveyed, 5: Very Well Conveyed
    \end{tcolorbox}
\caption{Questionnaire to evaluate elements relevancy, narrative coherence, and
educational value of the \texttt{\textbf{Tinker Tale}} generated stories.}
\label{fig:questionnaire}
\end{figure}

\subsection{Dialogue Manager}
\label{subsec:dialogue_manager}
Table \ref{tab:prompt} illustrates how the Dialogue Manager uses a pointer to control the game's progression and how the Prompt Converter converts the current input into the appropriate prompt.

\begin{table*}[t!]
\renewcommand{\arraystretch}{1.5} 
\centering
\scriptsize
\begin{tabular}{|p{0.2\linewidth}|p{0.75\linewidth}|} 
\hline
\multicolumn{1}{|c|}{\textbf{Pointer}} & \multicolumn{1}{c|}{\textbf{Prompt}} \\ \hline
\centering \texttt{Initiate} & You are a friendly intelligent agent named Tinker Tales, helping a child aged 4-6 with story generation using a game board. The child will communicate with you by scanning pawns or tokens or speaking, progressing through the game. The activity involves selecting three characters, then following the structure of introduction, development, crisis, and conclusion to create the story. First, the child will scan the three characters pawns. Then, using those characters, the child will complete the story by moving through the steps. For each part—introduction, development, crisis, and conclusion—the child will choose a place, an item, and a emotion. With the given three characters, place, item, and emotion, you will generate the story for each part. I will provide instructions at each step. To begin, introduce yourself to the child and tell them that the first thing to do is choosing characters they like and start creating the story with. Ask the child to scan the first character pawn for now. Your response will be delivered to the child via Text to Speech. \\ \hline
\centering \texttt{Character:1:Select} & The child chose \textcolor{blue}{\texttt{\$\{NFC\_WRITTEN\_TEXT\}}} as the first character. Ask the child to press the speaker button and say the name they want to give, along with describing the character's traits. \\ \hline
\centering \texttt{Character:1:Define} & The child said, "\textcolor{blue}{\texttt{\$\{SPEECH\_TO\_TEXT\}}}" Respond with empathy and tell the child to now scan the second character pawn. \\ \hline
\centering . . . & \\ \hline
\centering \texttt{Character:3:Select} & The child chose \textcolor{blue}{\texttt{\$\{NFC\_WRITTEN\_TEXT\}}} as the third character. Ask the child to press the speaker button and say the name they want to give, along with describing the character's traits. \\ \hline
\centering \texttt{Character:3:Define} & The child said, "\textcolor{blue}{\texttt{\$\{SPEECH\_TO\_TEXT\}}}" Respond with empathy to the child and ask to unfold the game board. Then, ask if they are ready and tell them to answer verbally. \\ \hline
\centering \texttt{Introduction:Ready} & The child said, "\textcolor{blue}{\texttt{\$\{SPEECH\_TO\_TEXT\}}}" Respond with empathy to the child and ask them to put all the character pawns on the start side of the game board. Then, ask if they are ready and tell them to answer verbally. \\ \hline
\centering \texttt{Introduction:Start} & The child said, "\textcolor{blue}{\texttt{\$\{SPEECH\_TO\_TEXT\}}}" Respond appropriately to the child. From now on, once the child chooses the place, item, and emotion for the characters, you will generate the introduction part. Briefly explain to the child what to do on this page, and then ask the child to choose a place, item, and emotion token, and place them one by one on the start page. Once everything is placed, tell the child to scan the place token first. Emphasize that after all the tokens are placed, they should scan the place token first. \\ \hline
\centering \texttt{Introduction:Place:Select} & The child chose \textcolor{blue}{\texttt{\$\{NFC\_WRITTEN\_TEXT\}}} as the place. Respond accordingly. Then encourage the child to describe the place in more detail after pressing the speaker button. What does this place look like, and what is the atmosphere like? \\ \hline
\centering \texttt{Introduction:Place:Define} & The child said, "\textcolor{blue}{\texttt{\$\{SPEECH\_TO\_TEXT\}}}". Respond with empathy to the child. Then, ask the child to scan the item token they choose. \\ \hline
\centering \texttt{Introduction:Item:Select} & The child chose \textcolor{blue}{\texttt{\$\{NFC\_WRITTEN\_TEXT\}}} as the item. Respond accordingly. Then encourage the child to describe the item in more detail after pressing the speaker button. Who brought this item? Or was it newly discovered in this place? \\ \hline
\centering \texttt{Introduction:Item:Define} & The child said, "\textcolor{blue}{\texttt{\$\{SPEECH\_TO\_TEXT\}}}". Respond with empathy to the child. Then, ask the child to scan the emotion token they choose. \\ \hline
\centering \texttt{Introduction:Emotion:Select} & The child chose \textcolor{blue}{\texttt{\$\{NFC\_WRITTEN\_TEXT\}}} as the emotion. Respond accordingly. Then encourage the child to describe the emotion in more detail after pressing the speaker button. Which character is feeling this emotion? Is there a reason for it? \\ \hline
\centering \texttt{Introduction:Emotion:Define} & The child said, "\textcolor{blue}{\texttt{\$\{SPEECH\_TO\_TEXT\}}}". Now, generate the Introduction of the story based on the three characters, place, item, and emotion. First, respond appropriately to the child and ask them to respond after pressing the speaker button if they would like to hear the Introduction part of the story. \\ \hline
\centering \texttt{Introduction:Complete} & The child said, "\textcolor{blue}{\texttt{\$\{SPEECH\_TO\_TEXT\}}}" First, create the story for the Introduction part, which should be about five to ten sentences long. Make sure the place, item, and emotion chosen by the child are naturally incorporated into the plot. Use simple words and sentence structures, keeping in mind that the listener is a child aged 4-6. Remember, the child should learn one of the following lessons by the end of the entire story, which includes the introduction, development, crisis, and conclusion. The lesson should be naturally woven into the story as it unfolds—there's no need to explicitly state it; just telling the story is enough. Lessons: [\textcolor{blue}{\texttt{\$\{LESSONS.join(", ")\}}}]. The key point is that it should connect naturally with the story you've created so far. The story should flow smoothly, with a logical and coherent progression, and connect well without any sudden or abrupt shifts. You need to explain the transitions between places naturally, ensuring the scenes follow one another in a way that feels seamless and cohesive. Here's the story you've made so far: \textcolor{blue}{\texttt{GENERATED\_STORY\_SO\_FAR}}. The next part can take place in a completely different location, so the events happening in the place of this part should be completed here. After telling the story, ask if the child liked it and encourage them to respond verbally. \\ \hline
\centering \texttt{Development:Ready} & The child said, "\textcolor{blue}{\texttt{\$\{SPEECH\_TO\_TEXT\}}}" Respond with empathy to the child and ask them to move the three character pawns to the journey side of the board. Then ask if they have moved them all and tell them to respond verbally. \\ \hline
\centering . . . & \\ \hline
\centering \texttt{Crisis:Ready} & The child said, "\textcolor{blue}{\texttt{\$\{SPEECH\_TO\_TEXT\}}}" Respond with empathy to the child and ask them to move the three character pawns to the climax side of the board. Then ask if they have moved them all and tell them to respond verbally. \\ \hline
\centering . . . & \\ \hline
\centering \texttt{Conclusion:Ready} & The child said, "\textcolor{blue}{\texttt{\$\{SPEECH\_TO\_TEXT\}}}" Respond with empathy to the child and ask them to move the three character pawns to the end side of the board. Then ask if they have moved them all and tell them to respond verbally. \\ \hline
\centering . . . & \\ \hline
\centering \texttt{Finish} & The child said, "\textcolor{blue}{\texttt{\$\{SPEECH\_TO\_TEXT\}}}" Respond with empathy to the child and say that the story is complete. Then, ask if the child want to hear the story from the beginning again. \\ \hline
\centering \texttt{Replay} & The child said, "\textcolor{blue}{\texttt{\$\{SPEECH\_TO\_TEXT\}}}" If the child wants, tell the entire story. After telling the story, kindly explain how to save this story. To save the story, simply say they want to save it and provide the title of the story. They can listen to the full story again anytime. \\ \hline
\end{tabular}
\caption{The Dialogue Manager controls the game progression using a pointer, while the Prompt Converter transforms the current input into the appropriate prompt based on the pointer's position. There are four types of dynamic inputs for the prompts: \textcolor{blue}{\texttt{LESSONS}} refers to a list of one-line lessons pre-entered by the parents, \textcolor{blue}{\texttt{NFC\_WRITTEN\_TEXT}} is the text written on the NFC chip, \textcolor{blue}{\texttt{SPEECH\_TO\_TEXT}} is the result of converting the child's speech into text, and \textcolor{blue}{\texttt{GENERATED\_STORY\_SO\_FAR}} refers to the story completed so far during the game progress.}
\label{tab:prompt}
\end{table*}

\end{document}